# Performance Level Evaluation Model based on ELM

Qian Mei
*State Key Laboratory of Airline Intergration Technology and Flight Simulation*
*COMAC Shanghai Aircraft Desin&Research Institute*
Shanghai, China
meiqian@comac.cc

*Abstract*—Human factor evaluation is crucial in designing civil aircraft cockpits. This process relies on the physiological and cognitive characteristics of the flight crew to ensure that the cockpit design aligns with their capabilities and enhances flight safety. Modern physiological data acquisition and analysis technology, developed to replace traditional subjective human evaluation, has become an effective method for verifying and evaluating cockpit human factors design. Given the high-dimensional and complex nature of pilot physiological signals, these uncertainties significantly impact pilot performance. This paper proposes a pilot performance evaluation model based on Extreme Learning Machine (ELM) to predict flight performance through pilots' physiological signals and further explores the quantitative relationship between human factors and civil aviation safety.

*Keywords—Human factor evaluation; pilot performance; Extreme Learning Machine (ELM); civil aviation safety*

## I. Introduction

The traditional method of civil aircraft cockpit human factors evaluation is mainly subjective evaluation. The subjective evaluation method has been used and iterated in the industry for many years, and has been proved to have good reliability and validity. However, due to the pursuit of research objectivity and system reliability, as well as the development of science and technology such as sensors, the collection and analysis of crew physiological data has gradually become an important method for cockpit human factor design verification. Physiological and psychological parameter signals of pilots are collected by various physiological measurement devices (including eye tracker, physiological meter, etc.) to study the relationship between physiological signals of pilots and flight performance, aiming to predict flight performance through physiological signals of pilots, and further explore the quantitative relationship and impact of human factors and civil aviation safety. In recent years, some literatures have shown that there is a certain relationship between pilots' multidimensional physiological parameters and flight performance, and the value and change of pilots' multidimensional physiological parameters will directly affect flight performance. L. Zhongqi realized flight performance evaluation based on BP neural network model by using real-time eye movement physiological data of pilots during flight [1]. PASSINO K M expounded the rationality of evaluating and predicting flight performance risk level based on multidimensional physiological signal parameters such as pilots' eye movement, heart rate and respiration [2]. In the research of risk assessment, Li Yunfei proposed an electric power risk load assessment model based on support vector machine (SVR). The model has good generalization ability, but its risk assessment ability is not very rational due to its slow convergence speed and other limitations [3]. Zhang Qingbao proposed a workload prediction model based on generalized neural network (GRNN), which has good prediction effect but its generalization is limited [4]. Yang Tingzhi proposed a research on power load prediction based on the combination of genetic algorithm and extreme learning machine (GA-ELM), but due to the limitations of genetic algorithm for initial population selection, the prediction results are difficult to achieve satisfactory results [5]. Based on the limitations of the above machine learning algorithms, this paper attempts to propose a flight performance risk assessment method based on BFA-ELM. Bacterial foraging algorithm (BFA) is a progressive evolutionary algorithm proposed by KM Passino in 2002 [6]. It is a global random search algorithm that combines the chemotaxis, population interaction, bacterial propagation and dispersion characteristics of the bacteria itself, and can effectively avoid local minima in the execution of the algorithm. These advantages effectively improve the prediction accuracy of the algorithm model. In recent years, BFA algorithm has also been applied in risk assessment and prediction. Extreme Learning Machine (ELM) is a single hidden layer feedforward neural network model based on generalized inverse matrix theory [7]. The ELM algorithm can operate very fast without the need for iteration of the hidden layer, and the input weight and bias can be arbitrarily set at the beginning of the algorithm, and the trained algorithm model can be obtained after the output weight is determined. This paper attempts to apply the bacterial foraging algorithm to the parameter optimization of ELM algorithm. The simulation results show that the algorithm has simple structure, fast operation speed, and better global searching ability and accuracy.

## II. Extreme Learning Machine (ELM) model

The theory of Extreme Learning Machine (ELM) was introduced by [8] in 2004. Specifically, for neural networks with a single hidden layer structure, the ELM model can randomly initialize the input weights and bias parameters and then compute the corresponding output weights. This process is significantly faster than the traditional Backpropagation (BP) neural network model due to ELM's rapid search speed. The algorithm structure of ELM is identical to that of a single-hidden layer feedforward neural network model. As illustrated in Fig.

1, the model contains only one hidden layer, making its structure simple and execution speed fast. Unlike the traditional BP neural network model, the ELM algorithm does not require iterative training of the hidden layers. Instead, it only needs to set the number of neurons in the hidden layer and then determine the output weights after training. The key advantage of ELM is its significantly higher training speed compared to traditional BP neural networks. Since its introduction in 2004, the ELM model has garnered considerable attention and application from many scholars due to its clear advantages in speed and efficiency.

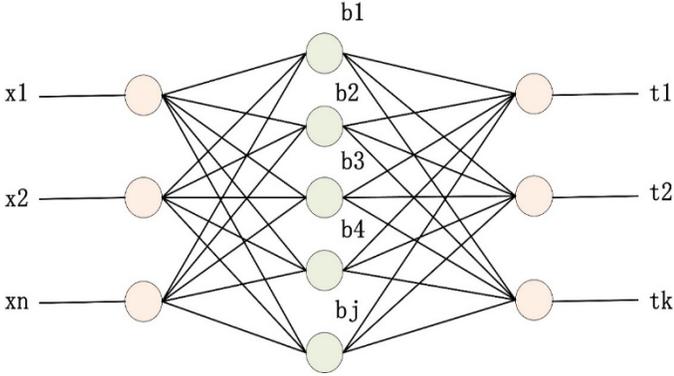

Fig. 1. Structure of extreme learning machine

The algorithm and training steps of ELM are described below:

Predict sample data set based on given training risk $D = \{(x_i, t_i) | x_i \in R^n, t_i \in R^n\}$, The activation function is f(x), and the number of hidden layer nodes is L. The learning goal of the single hidden layer neural network is to minimize the output error of the model, and the ELM regression model can be obtained:

$$\sum_{i=1}^{L} \beta_i f(a_i x_1 + b_i) = t_1$$

$$\sum_{i=1}^{L} \beta_i f(a_i x_2 + b_i) = t_2$$

$$\vdots \quad (1)$$

$$\sum_{i=1}^{L} \beta_i f(a_i x_k + b_i) = t_k$$

$a_i, i = 1, \cdots, L$ is the input weight, $b_i, i = 1, \cdots, L$ is the offset, k is the number of data samples. Formula (1) can be expressed as:

$$T_k = H_k \beta_k \quad (2)$$

$H_k$ is the output matrix of hidden layer neurons, expressed as:

$$H_k = \begin{bmatrix} f(a_1 x_1 + b_1) & \cdots & f(a_L x_1 + b_L) \\ \vdots & \ddots & \vdots \\ f(a_i x_k + b_i) & \cdots & f(a_L x_k + b_L) \end{bmatrix} \quad (3)$$

Formula (2), $\beta_k$ is the Output matrix weights, $T_k$ is the Output matrix, and from that we get Formula (4):

$$\beta_k = (H_k^T H_k)^{-1} H_k^T T_k \quad (4)$$

Based on the above derivation, the prediction model after training of ELM algorithm is obtained as follows:

$$y = \sum_{i=1}^{L} \beta_i f(a_i x + b_i) \quad (5)$$

Generally speaking, the training goal of the ELM algorithm model is to minimize the model's predictive training error, that is, to determine the output weight $\beta_k$, $a_i, i = 1, \cdots, L$ is the input weight, $b_i, i = 1, \cdots, L$ is the offset:

$$\|H(\hat{a}_i, \hat{b}_i)\hat{\beta}_i - T\| = min_{a,b,\beta} \|H(a_i, b_i)\beta_i - T\| \quad (6)$$

The above equation is equivalent to the training objective of the ELM algorithm model is to minimize the loss function:

$$E = \sum_{j=1}^{N} \left( \sum_{i=1}^{L} \beta_i f(a_i * X_j + b_i) - t_j \right)^2 \quad (7)$$

The training steps of ELM algorithm of extreme learning machine are summarized as follows:
1) Sample input: Input the risk prediction training sample data set, and test the remaining part of the sample data.
2) Dertermin the number of neuron nodes in the hidden layer, the activation activation function is f(x), and the parameters of each node in the hidden layer are initialized. $a_i, i = 1, \cdots, L$ is the input weight, $b_i, i = 1, \cdots, L$ is the offset.
3) Calculate the hidden layer neuron output matrix $H_k$.
4) Obtain the weight of the output matrix $\beta_k$ through the formula (4)
5) Finally, the prediction model results of ELM algorithm are obtained.

The optimization criterion of ELM is the minimization of the training error, and the output weights $\beta_k$ are obtained by assigning values to each parameter in the model. The core of ELM algorithm is to solve the output weight to minimize the error function. Compared with the traditional neural network algorithm, ELM has certain advantages in operation speed and prediction accuracy. However, in practical engineering applications, the ELM model may contain a large number of hidden layer nodes and the determination of $a_i$ and $b_i$ parameters in the model will affect the training effect of the model. These uncertainties make the ELM model algorithm have certain limitations in application.

III. FLIGHT PERFORMANCE RISK ASSESSMENT MODEL BASED ON BFA-ELM

A. Bacterial Feeding Algorithm (BFA)

Bacterial foraging algorithm is a new simulation algorithm proposed by K.M.Passino in 2002 based on the behavior of E. coli Ecoli devouring food in human intestines and simulating the foraging behavior of E. coli. In the bacterial foraging (BFA) algorithm model, there are three main parts in the algorithm structure: the bacterial chemotactic part, the bacterial reproduction part and the bacterial migration dispersal part. Here are some specific explanations for the three main sections.

Chemotactic stage: During the foraging process, E. coli will compare the current environmental state information with the previous state environment at a certain moment, and carry out a certain decision-making process to realize the bacteria's decision-making judgment feedback mechanism function on the environment. At this stage, bacteria will produce a series of optimization actions based on the judgment of the current environment, such as alternating forward, flipping, and stopping movement. The operator of the chemotactic phase can be expressed as:

$$X_i(j+1, k, l) = X_i(j, k, l) + R * s_p * \phi(i) \quad (8)$$

$$\phi(i) = \frac{X_i(j,k,l) - X_{rand}(j,k,l)}{\|X_i(j,k,l) - X_{rand}(j,k,l)\|} \quad (9)$$

In the formula, $X_i(j, k, l)$ represents the environmental position coordinates of the i bacteria during the J TH chemotactic operation, the k reproductive operation, and the l migratory operation. $X_{rand}(j, k, l)$ represents the domain environment range of $X_i(j, k, l)$. $s_p$ is the step length of the bacteria rolling in any direction in the current environmental domain, and R is a random variable with the value range of [0,1].

Reproductive replication stage: When the bacterial energy reaches a certain level, it completes the chemotactic behavioral operation and enters the next stage, the reproductive stage. The reproductive behavior of bacteria is similar to the survival of the fittest and survival of the fittest in nature, and only bacteria with high energy have the opportunity to reproduce, while bacteria with low energy are selected to enter the extinction stage. In the reproduction and replication stage of bacteria, the mother bacteria will multiply and split into two daughter bacteria, and the daughter bacteria will simulate the biological reproduction behavior and have the same step size, environmental location and other information as the mother bacteria.

Migration dispersal stage: Due to environmental changes such as environmental temperature and food conditions, the living area of bacterial individuals may change. Migration behavior is generated according to a certain probability of uncertainty. When migration behavior occurs, bacteria will produce new bacterial individuals at any location within the domain search range. Adding bacterial migration dispersal behavior to the bacterial foraging algorithm will help improve the global optimal search ability of the whole algorithm model.

The flowchart of the main steps of the bacterial foraging algorithm is shown in Fig. 2.

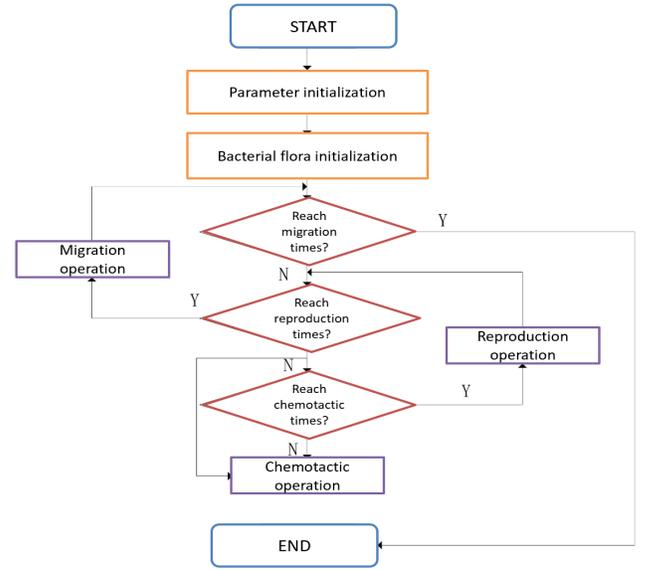

Fig. 2. Algorithm flow chart of bacteria foraging algorithm

B. Parameter optimization process of ELM model based on BFA algorithm

Multidimensional physiological parameters of pilots are nonlinear and uncertain. Input weight $a_i$, hidden layer bias $b_i$, and the number of hidden layer nodes L in the ELM algorithm model all have an impact on the predictive ability of the ELM flight performance model. Based on the superiority of BFA, this paper attempts to propose an ELM parameter optimization model based on BFA. The application of BFA in the parameter optimization of ELM model can make up for the limitation that the determination of $a_i$ and $b_i$ parameters in ELM model will affect the training effect of the model. The parameter optimization process of BFA-ELM is as follows:

1) Record and collect experimental data, and conduct unified and normalized processing of the data. The normalized calculation formula can be expressed as:

$$x_{norm} = \frac{x - x_{min}}{x_{max} - x_{min}} \quad (10)$$

In the formula, $x_{norm}$ is the data result after normalization processing, $x$ is the pilot physiological signal data that needs to be processed, that is, one of the five types of physiological signals. $x_{min}$ is the minimum value of this kind of pilot physiological signal data set, and $x_{max}$ is the maximum value of this kind of pilot physiological signal data set.

2) First, the initial values of each parameter in the bacterial foraging algorithm BFA were set. The number of bacteria S and the frequency distribution of chemotaxis, reproduction and replication of bacteria in the three stages, and migration and dispersion of bacteria corresponded to $N_c, N_{re}, N_{ed}$.

3) The input weight $a_i$, hidden layer bias $b_i$, and the number of hidden layer nodes L in the ELM algorithm model need to be optimized. The initial location of bacteria is assigned X according to the numerical region in these ELM parameters, and the value range of X is [0,1]. By comparing the predicted value of the flight performance model with the flight performance level, the prediction accuracy of the risk assessment model can be obtained, that is, the fitness $J_0$ of each bacteria.

4) $X_i(j, k, l)$ represents the relative environmental position of the i bacteria during the J TH chemotactic operation, the KTH reproductive replication operation, and the l migratory dispersal operation. In step (3), we have assigned the location of bacteria in the solution space of each parameter, and according to the model prediction accuracy obtained by bacteria in each spatial position, the fitness $J_i$ of bacteria in each available spatial position is obtained. If $J_i$ has a global minimum $J_{min}$ in the range of available positions in the solution space of the parameter, then $J_{min}$ is the optimal solution of the parameter, that is, the optimal fitness. Repeat the above steps to complete the number of stages set in step (2).

5) The optimal fitness of each bacterium in the BFA algorithm is the optimal solution of the corresponding parameters in the ELM algorithm. After optimizing the parameter setting of ELM based on BFA, the influence of the uncertainty of parameter selection in the ELM algorithm on the prediction and evaluation effect of the model itself can be optimized. Further, the flight performance risk assessment model based on BFA-ELM can be obtained.

## IV. EXPERIMENTAL SIMULATION AND RESULT ANALYSIS

### A. Flight performance risk assessment simulation experiment process

The flight crew is made up of 10 male pilots with an average age of 42 and an average flying time of nearly 6,000 hours. The flight simulator is an Airbus A320 flight simulator, which records basic flight data through the flight recorder during flight tasks, and the simulator sampling frequency is 30HZ. In the experiment, the pilot will also wear wearable devices to monitor the pilot's physiological parameters in real time, mainly including eye tracking recorder and ECG recorder. The eye tracker worn by the pilot can sample the pilot's pupil diameter and other data during flight, and the frequency of the eye tracker is 30HZ. The main objective of the ECG recorder is to sample the pilot's heart rate, breathing and other psychological state parameters in the experiment, and the sampling frequency of the ECG recorder is 1HZ. The task scenario in the experiment was the final approach and landing procedure. In the experiment, 10 pilots were divided into two roles: lead pilot and co-pilot. However, this paper only studied the influence of physiological parameters of the lead pilot on flight performance during each flight, and the influence of human factors of the co-pilot on flight performance was not considered in this paper. The data set of flight results of 20 flights were selected for statistical analysis.

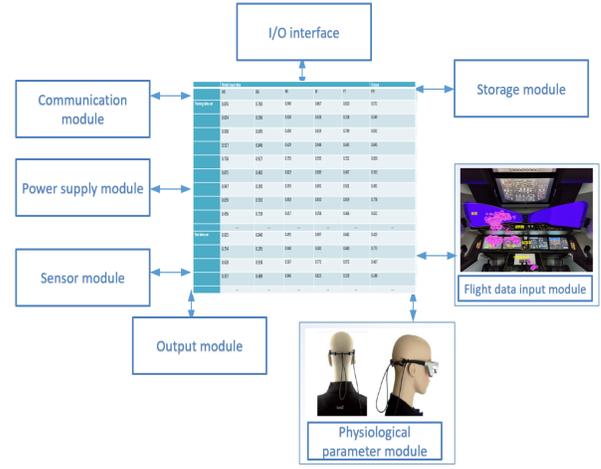

Fig. 3. Human factor quantification and flight performance evaluation model

In the collected pilot physiological signals, this paper selects the following five categories of pilot physiological signals: The data sets of five kinds of physiological signals, namely heart rate (HR), respiration amplitude (RA),respiration rate (RR), pilot blink interval (BI) and instrument fixation time (FT), are used as inputs of the flight performance risk assessment model based on BPA-ELM, and the corresponding flight performance level is used as outputs of the model [9]. In this paper, flight performance is defined as the mean square error of flight altitude during flight, and the corresponding flight performance index (FPI) level is defined as:

$$FPI = \frac{\sqrt{\sum_{i=1}^{N}(h_{ac}-h_{ex})^2}}{N} \quad (11)$$

In the formula, $h_{ac}$ is the actual flight altitude of the aircraft at each sampling point, $h_{ex}$ is the expected flight plan altitude of the aircraft at the sampling point, and N is the number of samples taken in a total period of time.

Among the 20 groups of flight data collected, 15 groups are selected as training sample data, and the remaining 5 groups of flight data are test sample data. After the normalization of all data, the input and output sample data sets for part of the training and testing in the experiment are listed in Table I.

TABLE I. THE DATA STYLE OF INPUT AND OUTPUT

|  | Model input data | | | | | Output |
|---|---|---|---|---|---|---|
|  | HR | RA | RR | BI | FT | FPI |
| Training data set | 0.876 | 0.783 | 0.949 | 0.867 | 0.810 | 0.571 |
|  | 0.654 | 0.596 | 0.928 | 0.928 | 0.318 | 0.549 |
|  | 0.938 | 0.695 | 0.450 | 0.619 | 0.749 | 0.832 |
|  | 0.517 | 0.846 | 0.629 | 0.948 | 0.645 | 0.695 |
|  | 0.758 | 0.927 | 0.793 | 0.593 | 0.352 | 0.650 |
|  | 0.875 | 0.482 | 0.829 | 0.589 | 0.407 | 0.763 |
|  | 0.947 | 0.395 | 0.593 | 0.691 | 0.501 | 0.491 |
|  | 0.659 | 0.592 | 0.818 | 0.810 | 0.819 | 0.778 |
|  | 0.456 | 0.739 | 0.617 | 0.358 | 0.466 | 0.652 |
|  | … | … | … | … | … | … |
| Test data set | 0.925 | 0.840 | 0.491 | 0.497 | 0.682 | 0.425 |
|  | 0.754 | 0.295 | 0.940 | 0.381 | 0.840 | 0.773 |
|  | 0.628 | 0.938 | 0.567 | 0.771 | 0.972 | 0.467 |

| 0.357 | 0.489 | 0.846 | 0.823 | 0.159 | 0.248 |
| ... | ... | ... | ... | ... | ... |

In the experiment, the accuracy of the model was verified by comparing the predicted value of the model for performance risk assessment with the real FPI value of the flight performance index level. At the same time, the prediction ability of the prediction model based on BPA-ELM is compared with that of the ELM model without parameter optimization of the BFA algorithm, so as to verify the effectiveness of the proposed BPA-ELM for flight performance risk assessment.

B. Experimental results and verification analysis

Five types of pilot physiological signals selected in this paper: heart rate (HR), respiration amplitude (RA), respiration rate (RR), pilot blink interval (BI) and instrument fixation time (FT) are selected based on relevant literature. In order to demonstrate that the five types of signals selected are reasonable and effective to represent and predict flight performance, correlation tests between the five types of physiological signal data and flight performance data are carried out below. Based on Pearson correlation coefficient analysis, the correlation between the five types of signals and flight performance parameter FPI is shown in Fig.4. Correlation test results of the five types of physiological signals show that there is a good correlation between the pilot physiological signals selected in the model and flight performance. These five kinds of physiological signals is reasonable and effective for the prediction of flight performance risk assessment model. Moreover, multidimensional physiological signals of pilots can indeed affect flight performance, and the study on the relationship between these physiological signals and flight performance risk level is of great significance for improving aviation safety level[10].

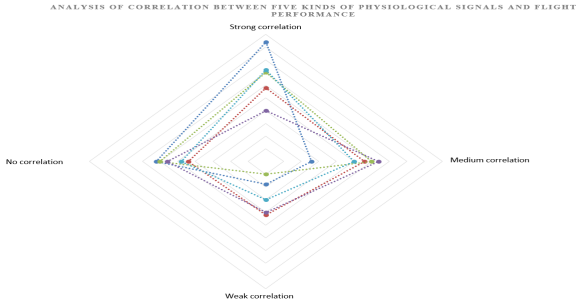

Fig. 4. Correlation analysis of five types of physiological signals with flight performance

The non-BFA optimized ELM model and the BFA-ELM optimized flight performance risk assessment model were used to predict the flight performance data of the same test set sample containing 100 data sets. The prediction results of the two algorithm models are shown in Fig.5 and Fig.6 respectively. It can be seen from the experimental results that the five physiological signals selected in this paper, namely heart rate (HR), respiration amplitude (RA), respiration rate (RR), pilot blink interval (BI) and instrument fixation time (FT), are reasonable for assessing flight performance risk level.

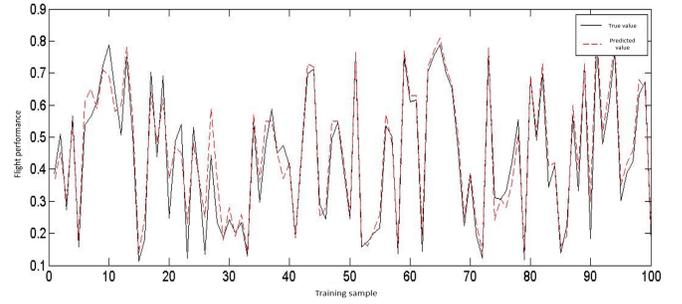

Fig. 5. Comparison between the predicted value and the true value of ELM algorithm without BFA optimization

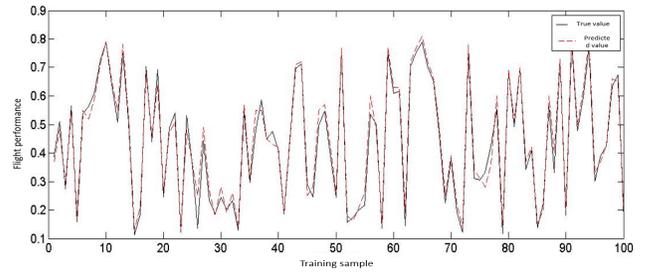

Fig. 6. Comparison between the predicted value and the true value of ELM algorithm based on BFA optimization

In order to further evaluate and analyze the prediction ability and effect of the two prediction models, MAE, MSE and MAPE were selected to evaluate the prediction ability and accuracy of the two prediction models. Table II lists the specific calculation formula and other information of these indicators parameters.

TABLE II. THREE EVALUATION INDICATORS

| Index parameter | Significance | Equation |
|---|---|---|
| MAE | Mean absolute error | $E_{MAE} = \sum_{i=1}^{n} |y_i - \hat{y}_i|/n$ |
| MSE | Mean square error | $E_{MSE} = \sum_{i=1}^{n} (y_i - \hat{y}_i)^2/n$ |
| MAPE | Mean absolute percentage error | $E_{MAPE} = \left(100 * \sum_{i=1}^{n} \left|\frac{y_i - \hat{y}_i}{y_i}\right|\right)/n$ |

In the table, $y_i$ represents the predicted flight performance level of the i th flight sample data model and $\hat{y}_i$ represents the true value of the flight performance level of the i th flight sample data. $n$ represents the number of flight sample datasets.

The calculation results of the three evaluation indicators are shown in Fig.7.

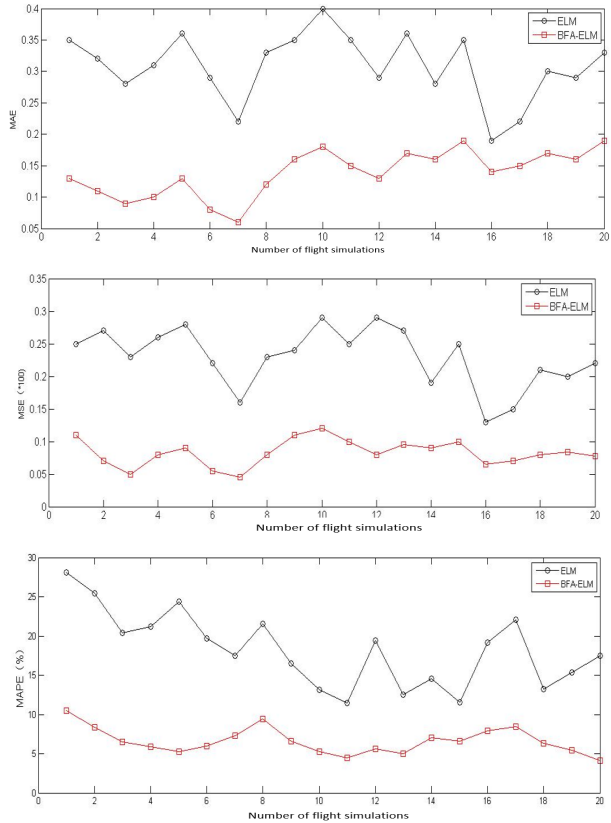

Fig. 7. Diagram of calculation results of three evaluation indexes

The experimental results show that under the three evaluation indexes, the prediction effect of the flight performance risk assessment prediction model proposed in this paper based on BFA optimization ELM parameter is better than that of the ELM model. The conclusion in this paper, based on the BFA - ELM flight performance risk assessment model is effective and can be further concluded in this paper, the pilot in the selection of model heart rate (HR), respiration amplitude (RA), respiration rate (RR), pilot blink interval (BI) and instrument fixation time (FT) of these five kinds of physiological signals is reasonable and effective for the prediction of flight performance risk assessment model. At the same time, through the calculation of evaluation indicators, it can be concluded that the average accuracy of the evaluation index of flight performance risk level based on BPA-ELM proposed in this paper can reach about 95%. The research content of this part on the relationship between multidimensional physiological parameters and flight performance of pilots has a good reference significance for the further improvement and accurate modeling of man-machine system model[11,12,13].

## V. Conclusion

Based on the flight performance risk assessment method of BPA-ELM, this paper firstly introduces the theoretical basis of Extreme Learning Machine (ELM) model and bacterial foraging algorithm (BFA). Then it introduces the process optimization algorithm of ELM model based on BFA algorithm to evaluate flight performance. The simulation experiment process of flight performance risk assessment was further introduced, and five physiological signals including pilot heart rate (HR), respiration amplitude (RA), respiration rate (RR), pilot blink interval (BI) and instrument fixation time (FT) were selected for flight performance risk assessment. Experimental results show that the ELM model algorithm based on BFA optimization established in this chapter is effective. n this paper, the study on the impact of multidimensional physiological signals of pilots on flight performance risk level complements the shortcomings of the current man-machine system modeling, which is difficult to characterize multidimensional physiological parameters of pilots. At the same time, it will also improve the man-machine system simulation model in the future, and provide a good reference for further accurate modeling.